# Impacts of COVID-19 control measures on tropospheric $NO_2$ over China, South Korea and Italy

Jiaqi Chen[1], Zhe Jiang[1], Kazuyuki Miyazaki[2], Rui Zhu[1], Xiaokang Chen[1], Chenggong Liao[3], Dylan B. A. Jones[4], Kevin Bowman[2], Takashi Sekiya[5]

[1]School of Earth and Space Sciences, University of Science and Technology of China, Hefei, Anhui, 230026, China.
[2]Jet Propulsion Laboratory, California Institute of Technology, Pasadena, CA, 91009, USA.
[3]Department of Oncology, Tangdu Hospital, Cancer Institute, Fourth Military Medical University, Xian, Shannxi, 710038, China.
[4]Department of Physics, University of Toronto, Toronto, ON, M5S 1A7, Canada.
[5]Japan Agency for Marine-Earth Science and Technology, Yokohama, 236-0001, Japan.

## Abstract

Tropospheric nitrogen dioxide ($NO_2$) concentrations are strongly affected by anthropogenic activities. Using space-based measurements of tropospheric $NO_2$, here we investigate the responses of tropospheric $NO_2$ to the 2019 novel coronavirus (COVID-19) over China, South Korea, and Italy. We find noticeable reductions of tropospheric $NO_2$ columns due to the COVID-19 controls by more than 40% over E. China, South Korea, and N. Italy. The 40% reductions of tropospheric $NO_2$ are coincident with intensive lockdown events as well as up to 20% reductions in anthropogenic nitrogen oxides ($NO_x$) emissions. The perturbations in tropospheric $NO_2$ diminished accompanied with the mitigation of COVID-19 pandemic, and finally disappeared within around 50-70 days after the starts of control measures over all three nations, providing indications for the start, maximum, and mitigation of intensive controls. This work exhibits significant influences of lockdown measures on atmospheric environment, highlighting the importance of satellite observations to monitor anthropogenic activity changes.

## Introduction

The COVID-19 has become a severe threat to global public health since it was initially reported in January 2020 (Zhu et al. 2020). The World Health Organization (WHO) declared COVID-19 as a global pandemic on Mar 11 2020, because of the rapid spread across the world: the reported confirmed cases are about 6400 thousand globally with 380 thousand deaths by June 1 2020 (http://www.chinacdc.cn). An important reason of the global outbreak of COVID-19 is lacking specific antiviral therapies and vaccines, and thus, the control strategy depends on isolation of cases and contact tracing to reduce the transmission rate (Chinazzi et al. 2020, Li et al. 2020), which has resulted in unprecedented lockdowns across the world.

As a precursor to ozone and secondary aerosols, $NO_2$ is one of the most important pollutants and plays a key role in tropospheric chemistry. Tropospheric $NO_2$ concentrations are strongly affected by fossil fuel combustions, such as power generation, industrial and transportation emissions (Jiang et al. 2018). The short lifetime of tropospheric $NO_2$ (few hours at the surface) makes it an ideal tracer for local anthropogenic emissions, as it exhibits marked responses to perturbations in economic activities (Mijling et al. 2009, Wang et al. 2015, Tong et al. 2016). The economic activity changes, due to the intensive lockdowns to mitigate the COVID-19, are expected to affect tropospheric $NO_2$ (Zhang et al. 2020), however, their actual influences are still uncertain, e.g., the "flawed estimates of the effects of lockdown measures on air quality derived from satellite observations" as suggested by the European Centre for Medium-Range Weather Forecasts (ECMWF 2020).

An important task of the international community, in 2020, is to understand the impacts of anthropogenic activity changes due to COVID-19 controls on atmospheric environment. In this work, we investigate the responses of tropospheric $NO_2$ to COVID-19 control measures over China, South Korea, and Italy to analyze the influence of lockdown measures on tropospheric $NO_2$, particularly, the responses of tropospheric $NO_2$ to the pandemic developments (i.e., start, maximum, and mitigation of pandemic spreads).

## Results

### Responses of tropospheric $NO_2$ to COVID-19

Figure 1a shows tropospheric $NO_2$ columns (OMI-QA4ECV, Boersma et al. 2018, See SI) over E. China, normalized in the 50-10 days before Jan 25 2020 (Spring Festival in 2020). The data over China are shifted for 2015-2019 to account for the economic cycles due to the Spring Festival. The reference time (RT, Table 1) is set to Jan 25 for the following two reasons: 1) the Spring Festival is a good indication for Chinese economic cycles; 2) tropospheric $NO_2$ in the 50-10 days before Jan 25 were not affected by COVID-19 (Figure 1b). Figures 1c-d and Figures 1e-f show tropospheric $NO_2$ and daily new confirmed cases over South Korea and N. Italy, respectively. The tropospheric $NO_2$ over South Korea is normalized in the 50-10 days before Feb 23 (RT, about 200 daily new confirmed cases). Considering the comparable populations between South Korea (about 50 million) and Italy (about 60 million), the tropospheric $NO_2$ over N. Italy is normalized in the 50-10 days before Feb 28 (RT, about 200 daily new confirmed cases) to ensure tropospheric $NO_2$ in the 50-10 days before the RTs were not affected by COVID-19 (Figures 1d, 1f).

As shown in Figure 1, the normalized tropospheric $NO_2$ changes (2020 vs. 2015-2019) exhibit the following relations with the COVID-19 pandemic developments:

1) Agreements in tropospheric $NO_2$ before the pandemic outbreaks: 50-0 days before the RT for E. China; 50-10 days before the RTs for South Korea and N. Italy.

2) Agreements in tropospheric $NO_2$ with pandemic mitigation: 60-80 days after the RT for E. China; 40-80 days after the RT for South Korea; 50-70 days after the RT for N. Italy.
3) Large differences in tropospheric $NO_2$ by more than 40%, coincident with the pandemic outbreaks.

Figure 2 shows the distributions of tropospheric OMI $NO_2$ columns over these three nations. Consistent with Figure 1, we find marked reductions of tropospheric $NO_2$ in the 10-30 days after the RTs over E. China, South Korea, and N. Italy in 2020. The reductions of tropospheric $NO_2$ are widely observable over these three nations.

Furthermore, the difference between tropospheric $NO_2$ in 2020 and 2015-2019 increased on the RT for E. China, but in about 10 days before the RTs for South Korea and N. Italy, suggesting a 10-day delay in the response of tropospheric $NO_2$ to the pandemic development in China compared to in South Korea and Italy. The delayed response in China could be due to the strong inhibition of the Spring Festival on Chinese economic activities, e.g., the E. China-averaged tropospheric $NO_2$ dropped by about 50% within 10 days prior to the national holiday (Figure 1a), which is even stronger than the perturbation due to COVID-19 controls. The perturbation in tropospheric $NO_2$ in the initial pandemic stage in China may have been covered by the inhibition due to the Spring Festival.

**Limited influences from non-anthropogenic processes**

We have demonstrated large perturbations in tropospheric $NO_2$ by more than 40% accompanied with the outbreaks of COVID-19. However, it is still unclear whether the perturbations were caused by anthropogenic or non-anthropogenic processes (e.g., large-scale anomaly in meteorological conditions). Figures 3a-c show modeled tropospheric $NO_2$ columns over these three nations, driven with the MIROC-Chem chemical transport model (See SI) and fixed anthropogenic $NO_x$ emissions in 2017. The meteorological fields are ERA-Interim with 1.125°x1.125° horizontal resolution. Considering the local equator crossing time (13:45) of OMI instrument, we only consider tropospheric $NO_2$ in 12:00-15:00 local time (Shen et al. 2019). The modeled tropospheric $NO_2$ are generally within the ±20% range of the 2015-2019 averages (shaded areas), with good agreement between 2020 (red) and 2015-2019 (blue).

Similarly, Figures 3d-f show modeled tropospheric $NO_2$ columns from the GEOS-Chem chemical transport model (See SI) and fixed anthropogenic $NO_x$ emissions in 2017. The meteorological fields are MERRA-2 with 2°x2.5° horizontal resolution. The modeled tropospheric $NO_2$ with GEOS-Chem are generally within the ±20% (E. China and South Korea) and ±30% (N. Italy) ranges of the 2015-2019 averages, with good agreement between 2020 and 2015-2019. Furthermore, Figures 3g-i show tropospheric OMI $NO_2$ columns. The observed tropospheric OMI $NO_2$ are generally within the ±20% range of the 2015-2019 averages, however, with significant discrepancy between 2020 and 2015-2019. The agreements between modeled and observed tropospheric $NO_2$ in Figure 3 suggest that the non-anthropogenic processes have limited influences on the observed $NO_2$ changes: about 20% for E. China and South Korea, and 20-30% for N. Italy, providing estimations for the uncertainties in the observed OMI $NO_2$ (Table 1).

The distributions of tropospheric OMI $NO_2$ in the 2015-2019 (Figures 3g-i) are shown in Figure 1 as the shaded areas. It demonstrates the deviations (larger than 40%) as well as the recovery of tropospheric $NO_2$ are caused by changes in anthropogenic $NO_x$ emissions. In addition, we find the trends in tropospheric $NO_2$ over N. Italy are almost the same in the 40-day period (RT to 40 days after the RT, Figure 1e) between 2015-2019 and 2020. It is thus, difficult to distinguish the changes in tropospheric $NO_2$ due to COVID-19 controls and climatological projections with simple comparison, as suggested by the European Centre for Medium-Range Weather Forecasts (ECMWF 2020).

**Impacts of lockdowns on atmospheric environment**

The above analysis indicates the important influences of anthropogenic activities on the observed tropospheric $NO_2$ changes. As shown in Table 1, the differences between tropospheric $NO_2$ in 2020 and 2015-2019 are larger than 40% in more than 17 days over E. China, South Korea, and N. Italy. Here we further investigate the relations between changes in tropospheric $NO_2$ and lockdown measures:

1) China: lockdowns in provinces outside of Hubei since around Jan 31, 2020 (#1, Wiki 2020). Considering the inhibition of the Spring Festival on Chinese economic activities, the lockdown measure (#1, Figure 1a) matches well with the start of the 40% perturbation in tropospheric $NO_2$.
2) South Korea: maximum quarantine in Gyeongsangbuk-Do (the province that COVID-19 was initially outbreak in South Korea) on Feb 25 2020 (#2, YNA 2020). As shown in Figure 1c, the quarantine (#2) matches well with the start of the 40% perturbation in tropospheric $NO_2$.
3) Italy: lockdown in N. Italy on Mar 7 2020 (#3, BBC 2020); all unnecessary commercial activities stopped on Mar 11 (#4, Repubblica 2020). As shown in Figure 1e, these lockdown measures (#3 and #4) match well with the start of the 40% perturbation in tropospheric $NO_2$.

The coincidences among the intensive lockdown measures, the 40% perturbations in tropospheric $NO_2$ and the mitigation of the pandemic demonstrate the influences of lockdown measures on atmospheric environment and pandemic developments. The recovery of tropospheric $NO_2$ around 40-60 days after the RTs provides indications for the mitigation of intensive controls.

Finally, we evaluate the impacts of lockdown measures on anthropogenic $NO_x$ emissions. Following Miyazaki et al. (2020), we constrain anthropogenic $NO_x$ emissions with an ensemble Kalman Filter (EnKF) while improving the representation of the chemical system (e.g., $NO_x$ lifetime) by assimilating multiple chemical species (See SI). The combined total (anthropogenic, soil, and lightning) emission is optimized in data assimilation with 1.125°x1.125° horizontal resolution. As shown in Figure 4, we find:

1) Agreements in anthropogenic $NO_x$ emissions before the pandemic outbreaks: 50-0 days before the RTs for E. China, South Korea, and N. Italy.
2) Differences in anthropogenic $NO_x$ emissions by up to 20%, coincident with the pandemic outbreaks over all three nations.

The similar responses of OMI $NO_2$ and derived anthropogenic $NO_x$ emissions to the pandemic outbreaks provide support to our conclusion. The relative uncertainties in the derived $NO_x$ emissions are larger than those in OMI $NO_2$. It could be partially associated with the region-specific data filters (Figure S1, See SI), which were not considered in the global assimilation. In addition, the perturbations in OMI $NO_2$ will become about 30% without the region-specific data filters (Figure S2, See SI), implying a ratio of 0.7 between changes in $NO_x$ emissions and tropospheric $NO_2$ columns, consistent with the reported non-linear relationship (Lamsal et al. 2011; Gu et al. 2016).

Using space-based measurements, this work exhibits important impacts of COVID-19 control measures on atmospheric environment: tropospheric $NO_2$ columns were reduced by 40%, and with up to 20% reductions in anthropogenic $NO_x$ emissions over E. China, South Korea, and N. Italy. More efforts are required to better understanding the worldwide responses of primary atmospheric pollutants to the lockdown measures, as they provide important information for the impacts of anthropogenic activities on atmospheric environment. In addition, the satellite data provides indications for the start (about 10 days before the RTs), maximum (about 10-20 days after the RTs), and mitigation (about 40-60 days after the RTs) of intensive COVID-19 controls, highlighting the importance of satellite observations, as a powerful tool, to monitor anthropogenic activity changes.

**Acknowledgments:** We acknowledge useful discussions with Folkert Boersma. We thank the National Health Commission (NHC) and the World Health Organization (WHO) for providing the COVID-19 daily new confirmed case data. We thank the providers of the OMI tropospheric $NO_2$ column data. This work is funded by the Hundred Talents Program of Chinese Academy of Science, and the Fundamental Research Funds for the Central Universities. The numerical calculations in this paper have been done on the supercomputing system in the Supercomputing Center of University of Science and Technology of China. Part of the research was carried out at the Jet Propulsion Laboratory, California Institute of Technology, under a contract with the National Aeronautics and Space Administration.

**Legends of Figures and Tables**

**Figure 1.** (A,C,E) Tropospheric OMI $NO_2$ columns (averaged in the period of ± 7 days with unit 1e15 molec/cm2) in 2020 and 2015-2019, normalized in the 50-10 days before the reference times (RT, magenta lines): Jan 25 (China), Feb 23 (South Korea) and Feb 28 (Italy). The shaded areas represent distributions of OMI $NO_2$ in 2015-2019. The shadow (green) shows the days with perturbations in tropospheric OMI $NO_2$ larger than 40%. The arrows show the events of COVID-19 controls. (B,D,F) Numbers of daily new confirmed cases of COVID-19. The jump on Feb 12 2020 (panel b) was caused by the change of testing methods, by reporting the cumulative clinically diagnosed patients as daily new confirmed cases (See SI). The E. China domain (land only) is defined by Figure 2a. The areas outside of China are excluded in the E. China domain. The N. Italy domain is defined as the north of 43°N (land only).

**Figure 2.** Tropospheric OMI $NO_2$ columns with unit 1e15 molec/cm2. (A,E,I) averages in the 40-20 days (before the RTs) in 2015-2019; (B,F,J) averages in the 10-30 days (after the RTs) in 2015-2019; (C,G,K) averages in the 40-20 days (before the RTs) in 2020; (D,H,L) averages in the 10-30 days (after the RTs) in 2020.

**Figure 3.** (A-C) Tropospheric $NO_2$ columns from MIROC-Chem model (12-15 local time, averaged in the period of ± 7 days with unit 1e15 molec/cm2) in 2015-2020, normalized in the 50-10 days before the reference times (magenta lines): Jan 25 (China), Feb 23 (Korea) and Feb 28 (Italy). (D-F) Same as panels a-c, but for tropospheric $NO_2$ columns from GEOS-Chem model. (G-I) Same as panels a-c, but for tropospheric $NO_2$ columns from OMI. The anthropogenic emissions in MIROC-Chem and GEOS-Chem models are fixed in 2017. The shaded areas show the ranges of ±20% of the 2015-2019 averages (±30% in panel f).

**Figure 4.** Derived anthropogenic $NO_x$ emissions (averaged in the period of ± 7 days with unit 1e-11 kgN/m2/s) in 2020 and 2015-2019, normalized in the 50-10 days before the reference times (magenta lines): Jan 25 (China), Feb 23 (South Korea) and Feb 28 (Italy). The shaded areas represent distributions of the derived $NO_x$ emissions in 2015-2019.

**Table 1.** The perturbations in OMI $NO_2$ are defined as the number of days with perturbations larger than 30%, 40% and 50%. The uncertainties in OMI $NO_2$ are defined based on the spreads in modeled and observed $NO_2$ (2015-2019, Figure 2). The perturbations in the derived $NO_x$

emissions are defined as the number of days with perturbations larger than 10%, and the maximum perturbations. The uncertainties in the derived $NO_x$ emissions are defined based on the spreads of the derived $NO_x$ emissions (2015-2019).

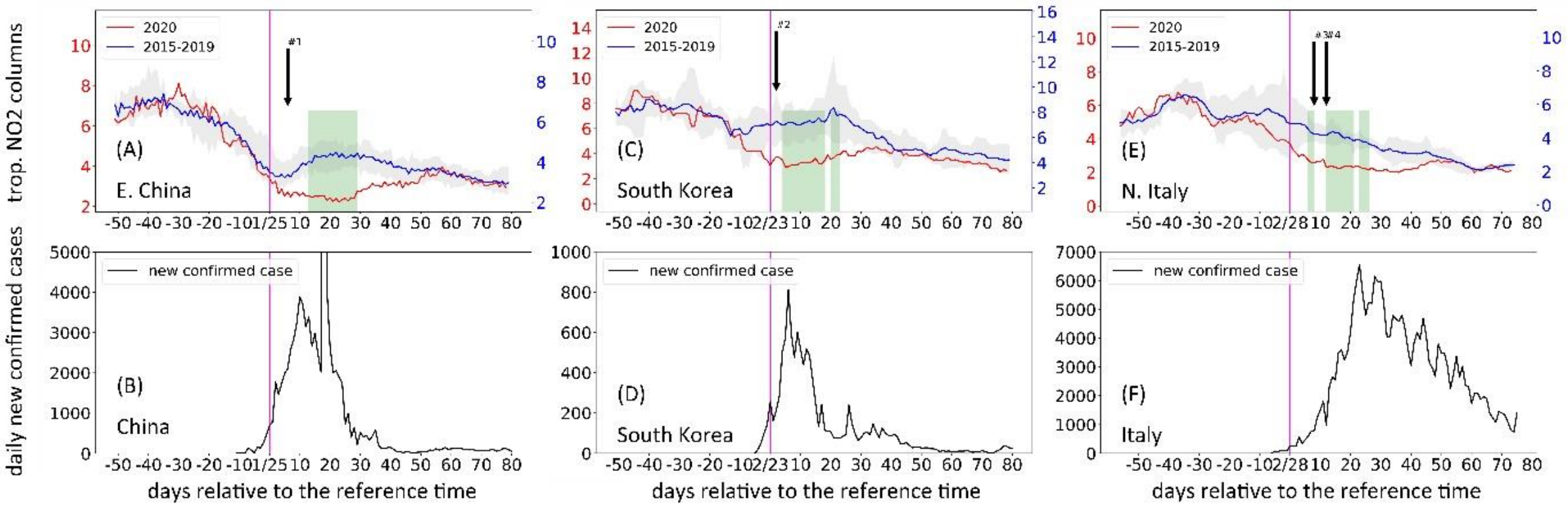

**Figure 1.** (A,C,E) Tropospheric OMI $NO_2$ columns (averaged in the period of $\pm$ 7 days with unit 1e15 molec/cm2) in 2020 and 2015-2019, normalized in the 50-10 days before the reference times (RT, magenta lines): Jan 25 (China), Feb 23 (South Korea) and Feb 28 (Italy). The shaded areas represent distributions of OMI $NO_2$ in 2015-2019. The shadow (green) shows the days with perturbations in tropospheric OMI $NO_2$ larger than 40%. The arrows show the events of COVID-19 controls. (B,D,F) Numbers of daily new confirmed cases of COVID-19. The jump on Feb 12 2020 (panel b) was caused by the change of testing methods, by reporting the cumulative clinically diagnosed patients as daily new confirmed cases (See SI). The E. China domain (land only) is defined by Figure 2a. The areas outside of China are excluded in the E. China domain. The N. Italy domain is defined as the north of 43°N (land only).

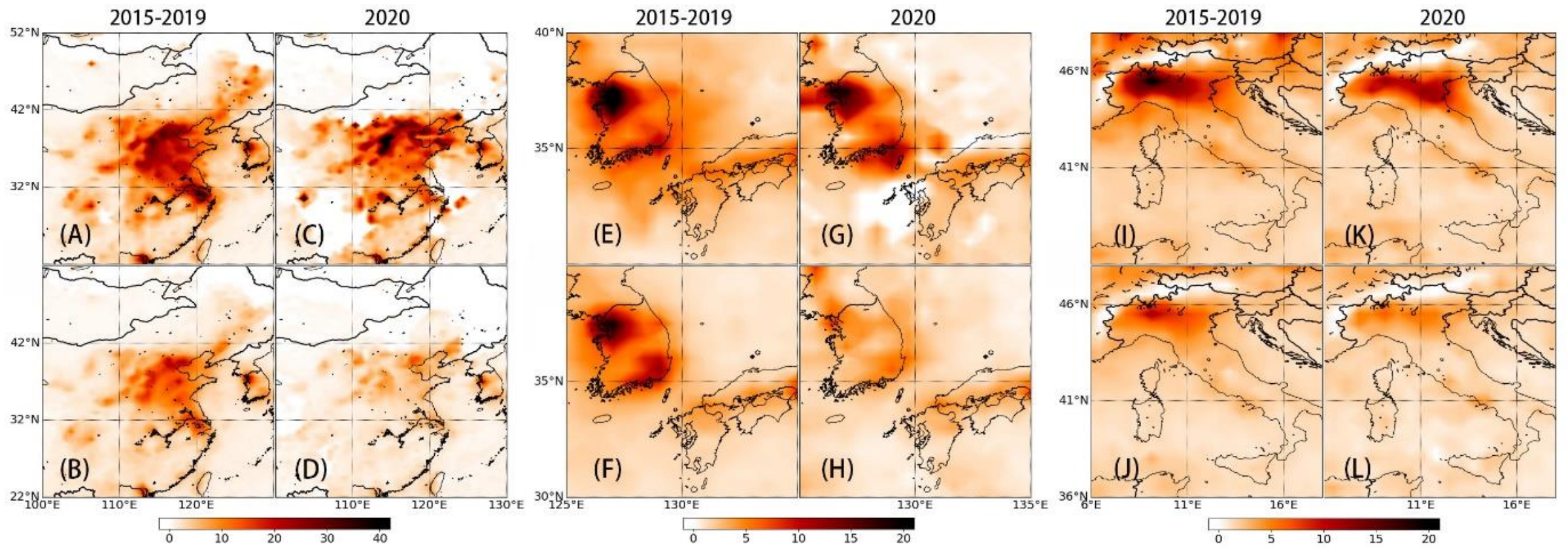

**Figure 2.** Tropospheric OMI $NO_2$ columns with unit 1e15 molec/cm2. (A,E,I) averages in the 40-20 days (before the RTs) in 2015-2019; (B,F,J) averages in the 10-30 days (after the RTs) in 2015-2019; (C,G,K) averages in the 40-20 days (before the RTs) in 2020; (D,H,L) averages in the 10-30 days (after the RTs) in 2020.

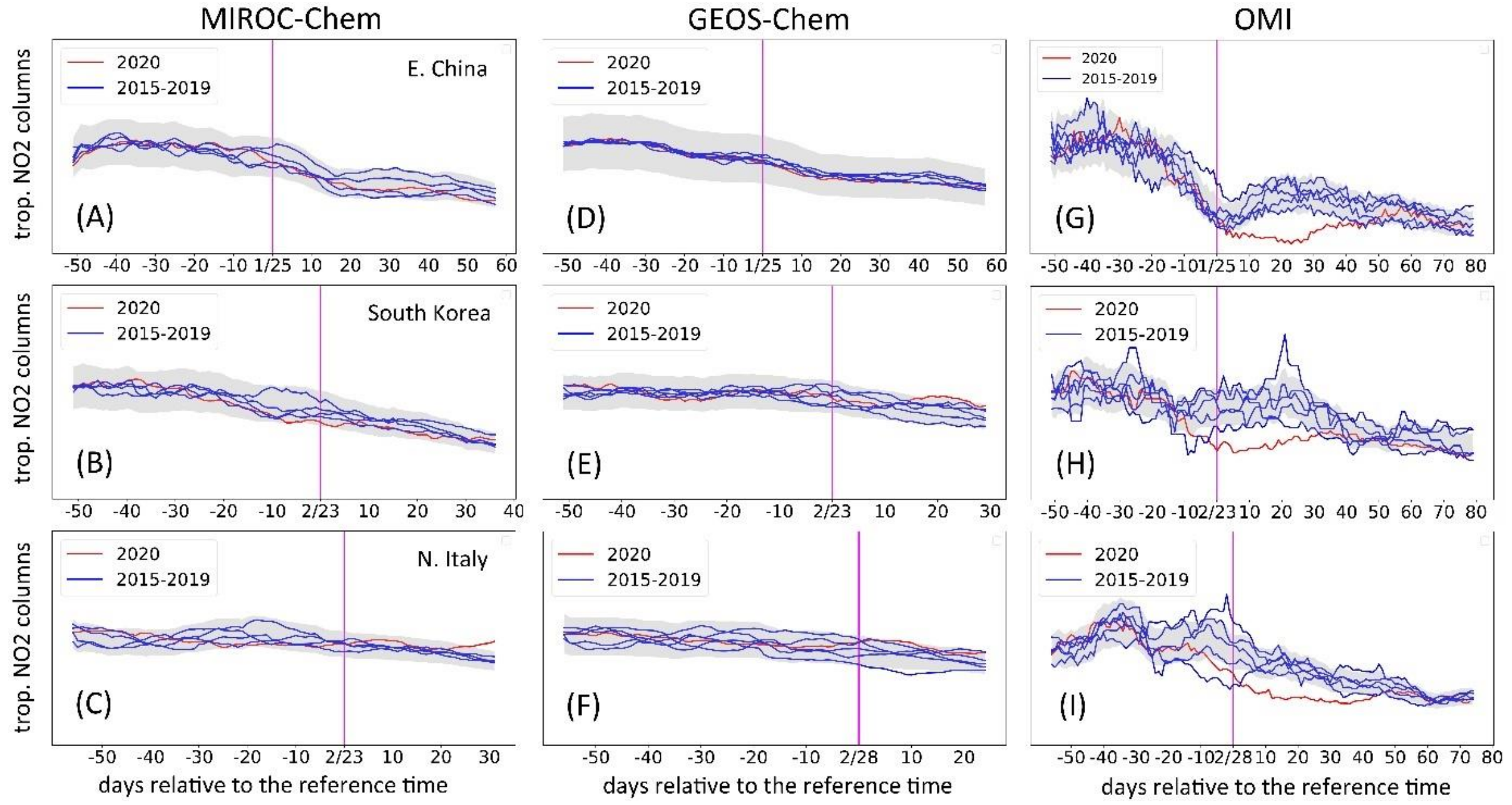

**Figure 3.** (A-C) Tropospheric $NO_2$ columns from MIROC-Chem model (12-15 local time, averaged in the period of ± 7 days with unit 1e15 molec/cm2) in 2015-2020, normalized in the 50-10 days before the reference times (magenta lines): Jan 25 (China), Feb 23 (Korea) and Feb 28 (Italy). (D-F) Same as panels a-c, but for tropospheric $NO_2$ columns from GEOS-Chem model. (G-I) Same as panels a-c, but for tropospheric $NO_2$ columns from OMI. The anthropogenic emissions in MIROC-Chem and GEOS-Chem models are fixed in 2017. The shaded areas show the ranges of ±20% of the 2015-2019 averages (±30% in panel f).

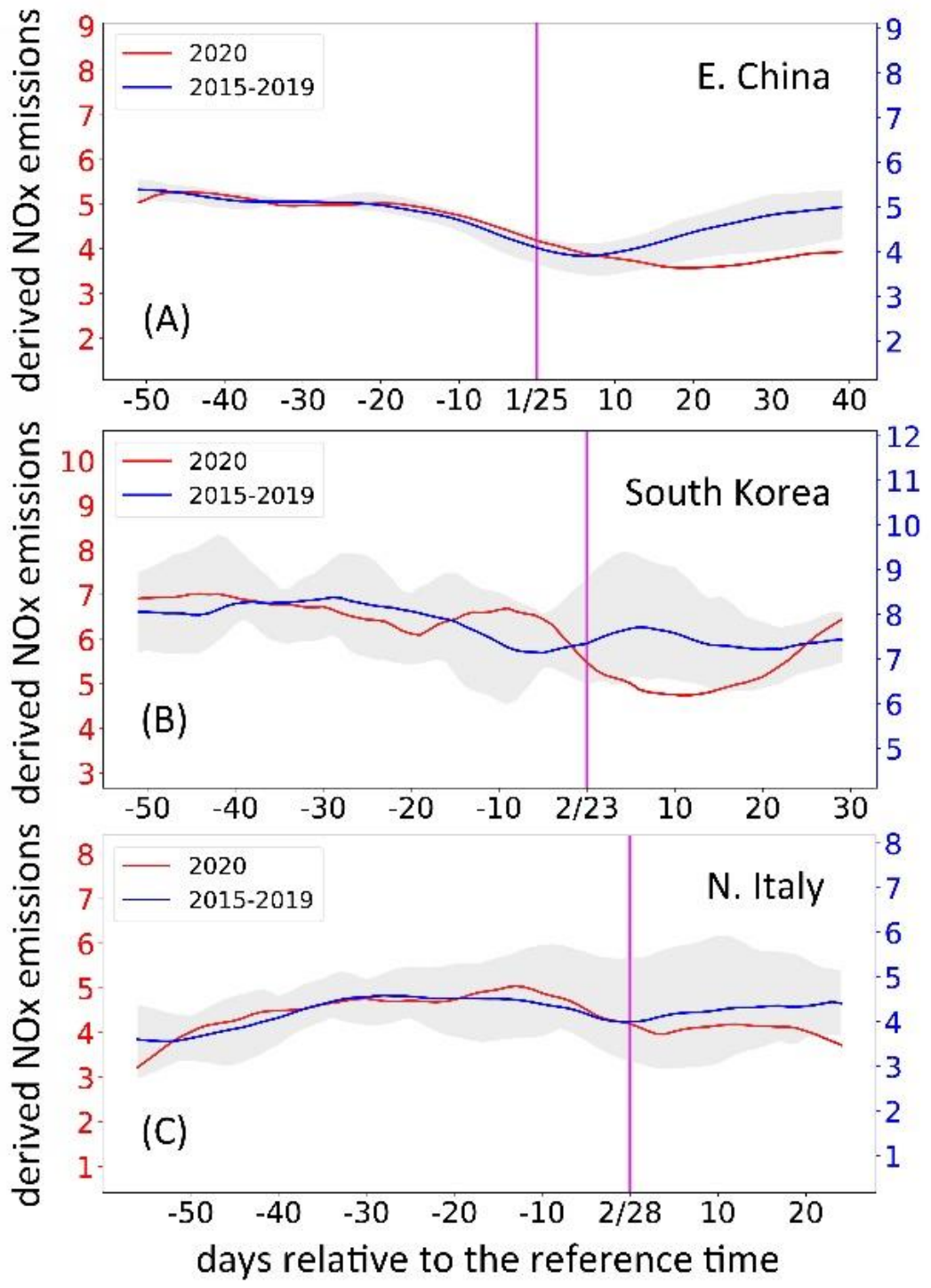

**Figure 4.** Derived anthropogenic $NO_x$ emissions (averaged in the period of ± 7 days with unit 1e-11 kgN/m2/s) in 2020 and 2015-2019, normalized in the 50-10 days before the reference

times (magenta lines): Jan 25 (China), Feb 23 (South Korea) and Feb 28 (Italy). The shaded areas represent distributions of the derived $NO_x$ emissions in 2015-2019.

| Domains | Reference time (RT) | Perturb. in OMI NO2 | | | | Perturb. in derived NOx | | |
|---|---|---|---|---|---|---|---|---|
| | | >30% (days) | >40% (days) | >50% (days) | uncertainties | >10% (days) | max | uncertainties |
| E. China | Jan 25 | 24 | 17 | 3 | ± 20% | 25 | 21% | ± 15% |
| South Korea | Feb 23 | 30 | 21 | 1 | ± 20% | 20 | 20% | ± 20% |
| N. Italy | Feb 28 | 40 | 20 | 0 | ± 20-30% | 10 | 21% | ± 30% |

**Table 1.** The perturbations in OMI $NO_2$ are defined as the number of days with perturbations larger than 30%, 40% and 50%. The uncertainties in OMI $NO_2$ are defined based on the spreads in modeled and observed $NO_2$ (2015-2019, Figure 2). The perturbations in the derived $NO_x$ emissions are defined as the number of days with perturbations larger than 10%, and the maximum perturbations. The uncertainties in the derived $NO_x$ emissions are defined based on the spreads of the derived $NO_x$ emissions (2015-2019).

# Supplemental Information

## Tropospheric OMI $NO_2$ column data

The OMI instrument on the Aura spacecraft has a spatial resolution of 13 km x 24 km (nadir view), which is in a sun-synchronous ascending polar orbit with a local equator crossing time of 13:45. OMI provides global coverage with measurements of both direct and atmosphere-backscattered sunlight in the ultraviolet-visible range from 270 to 500 nm; the spectral range 405-465 nm is used to retrieve tropospheric $NO_2$ columns. The OMI retrievals (level 2, QA4ECV, Boersma et al. 2018) are used in this work. Following Jiang et al. (2018), and the QA4ECV Product User Manual (http://www.qa4ecv.eu/ecv/no2-pre/data), the following filters are applied in our analysis:

1) Tropospheric Column Flag = 0
2) Surface Albedo < 0.3
3) Cloud Radiance Fraction < 0.5
4) No edge data (rows 1-5, 56-60)
5) No row anomaly data (rows 27-55 for the period 2015-2020)

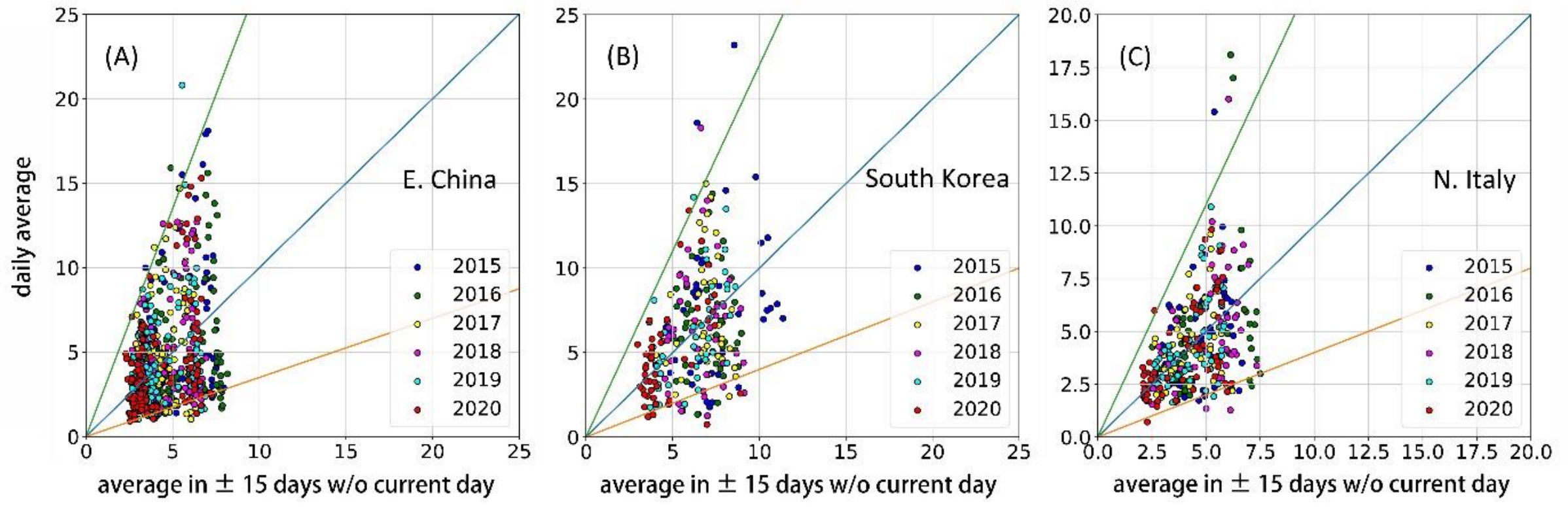

**Figure S1.** y-axis: regional daily average of tropospheric OMI $NO_2$ columns; x-axis: regional average of tropospheric OMI $NO_2$ columns in the period of ± 15 days by excluding the current day.

Besides the aforementioned filters, the regional averaged OMI $NO_2$ data are affected by the different daily coverage of satellite data. Figure S1 shows the relations between daily average of tropospheric OMI $NO_2$ and the average of its neighbouring days (± 15 days without the current day). Large deviation from the 1:1 relationship means the daily average of OMI $NO_2$ is pronounced higher (or lower) than its neighbouring days. The following region-specific filters (green lines in Figure S1) are supplemented in our analysis:

6) E. China: $0.35x < y < 2.7x$

7) Korea: $0.4x < y < 2.2x$

8) N. Italy: $0.4x < y < 2.2x$

Figures S2a-c show tropospheric OMI $NO_2$ columns in 2015-2020. The observed tropospheric OMI $NO_2$ are generally within the ±30% range of the 2015-2019 averages (shaded areas). The application of the region-specific quality filters reduced the random uncertainties from ±30% to ±20% (Figures S2d-f), while keeping the consistent patterns in the normalized $NO_2$ (Figures S2g-i).

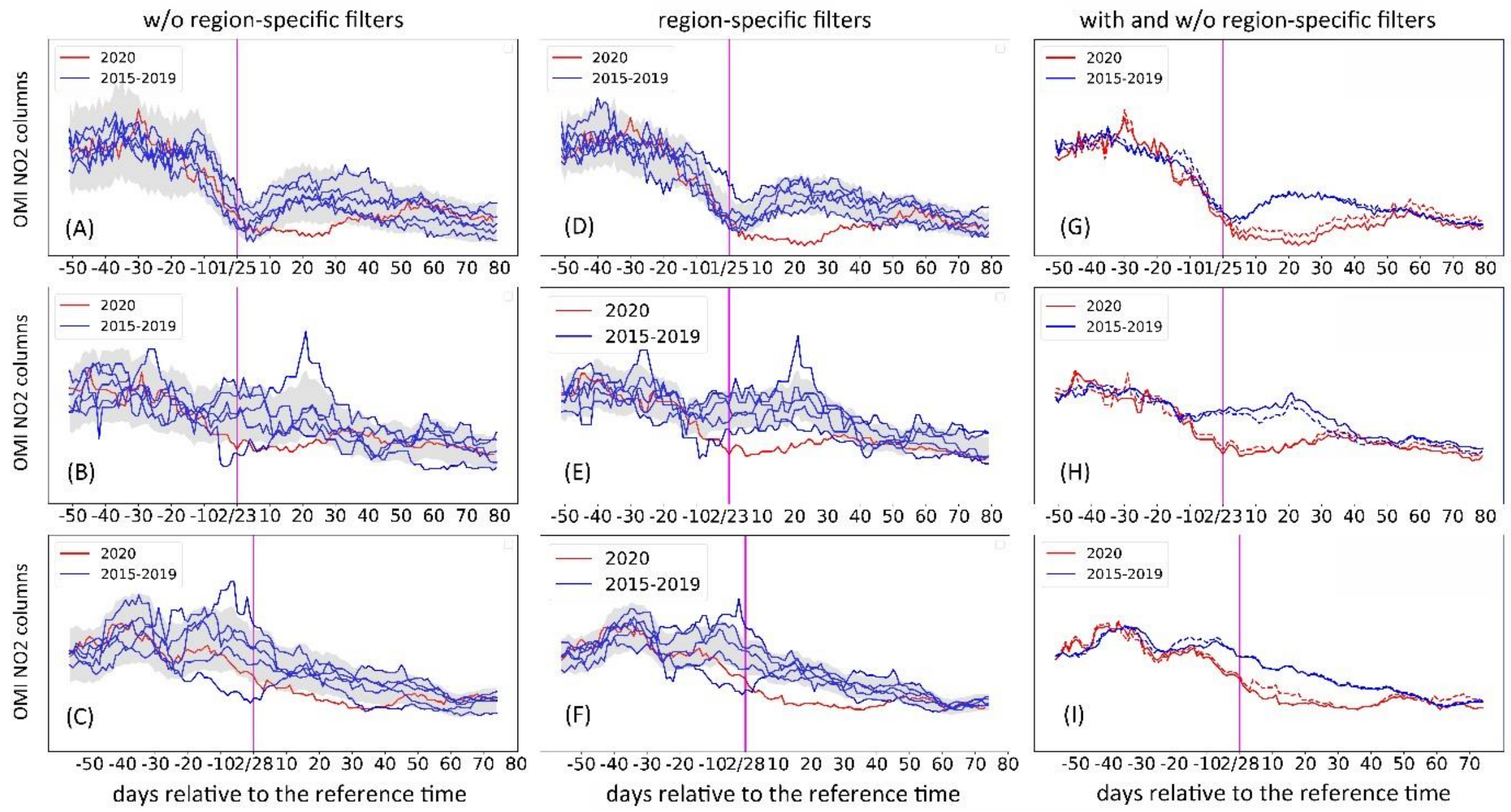

**Figure S2.** Tropospheric OMI $NO_2$ columns (averaged in the period of ± 7 days with unit 1e15 molec/cm2) in 2015-2020, normalized in the 50-10 days before the reference times (magenta lines): Jan 25 (China), Feb 23 (Korea) and Feb 28 (Italy). The shaded areas show the ranges of ±30% (panels a-c) and ±20% (panels d-f) of the 2015-2019 averages. The solid and dashed lines in panels e-i show tropospheric OMI $NO_2$ columns with and without the region-specific filters.

**MIROC-Chem model simulations:** The MIROC-Chem chemical transport model (Watanabe et al., 2011) with 1.125°x1.125° horizontal resolution for 2014-2020 are used in this work. The anthropogenic emissions were fixed in 2017. The model considers detailed photochemistry in the troposphere and stratosphere and is coupled to the atmospheric general circulation model MIROC-AGCM version 4 (Watanabe et al., 2011). The meteorological fields simulated by MIROC-AGCM were nudged toward the six-hourly ERA-Interim (Dee et al., 2011). The MIROC-Chem model has been widerly used in global atmospheric chemistry studies (Jiang et al. 2018; Miyazaki et al. 2017; Miyazaki et al. 2020).

**GEOS-Chem model simulations:** The GEOS-Chem chemical transport model (www.geos-chem.org, version 11) with 2°x2.5° horizontal resolution for 2014-2020 are used in this work. The anthropogenic emissions were fixed in 2017. The standard GEOS-Chem chemical mechanism includes 68 tracers, which can simulate detailed tropospheric $O_3$-$NO_x$-hydrocarbon chemistry, including the radiative and heterogeneous effects of aerosols. The model is driven by assimilated meteorological fields from the Modern-Era Retrospective analysis for Research and Applications, Version 2 (MERRA-2).

**Derived anthropogenic $NO_x$ emission estimates:** Based on an ensemble Kalman filter technique, Miyazaki et al. (2017) estimated global surface $NO_x$ emissions for the period of 2005-2015 by assimilating multiple satellite data sets. Using the OMI QA4ECV $NO_2$ products (Boersma et al. 2018), updated emission estimates with 1.125°x1.125° horizontal resolution for 2014-2020 are used in this work. The combined total (anthropogenic, soil, and lightning) emission is optimized in data assimilation. This is to avoid the difficulty associated with optimizing the spatiotemporal structure in background errors for each category source separately. In our analysis, individual emission sources were estimated using the emission ratio between different categories in the a priori emission inventories. The forecast model is MIROC-Chem (Watanabe et al., 2011).

**COVID-19 daily new confirmed case data:** The COVID-19 confirmed case data is downloaded at the Chinese Center for Disease Control and Prevention network (http://www.chinacdc.cn/), in which the data is provided by the National Health Commission (NHC) and the World Health Organization(WHO). As shown in Figure S3, the data from the NHC/WHO is consistent but smoother than the data from Johns Hopkins University (https://www.arcgis.com/apps/opsdashboard/index.html#/bda7594740fd40299423467b48e9ecf6). The NHC (Hubei province, China) changed the testing methods on Feb 12 2020 by considering patients who have been clinically diagnosed as COVID-19 disease as confirmed cases. The cumulative number of clinically diagnosed patients was reported as daily new confirmed cases on Feb 12 2020, which resulted in a jump by 14840.

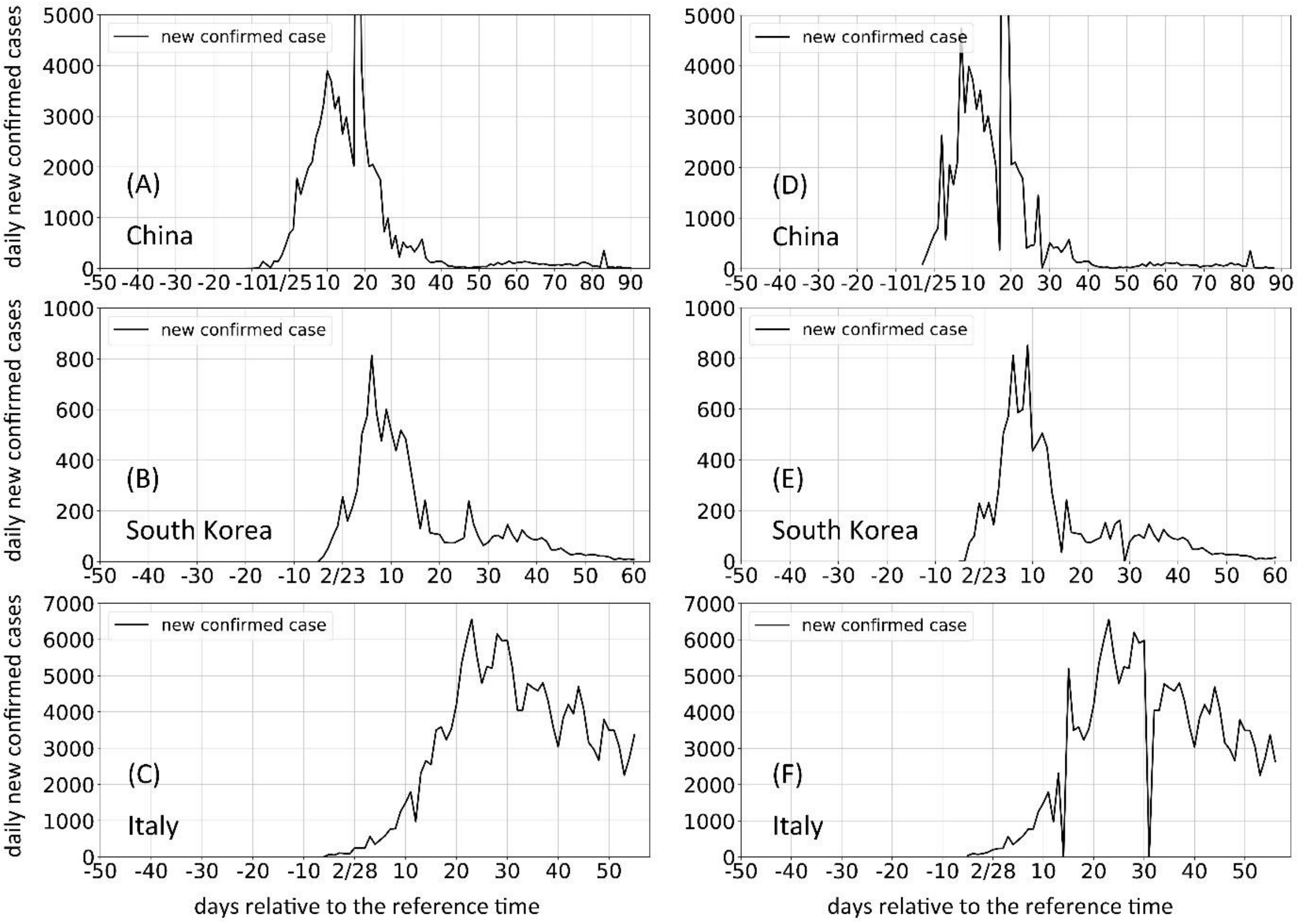

**Figure S3**. Daily new confirmed cases of COVID-19 from (A-C) National Health Commission and World Health Organization; (D-F) John Hopkins University.